Alexey Markov

# SOFTWARE TESTING MODELS AGAINST INFORMATION SECURITY REQUIREMENTS

*An overview and classification of software testing models are done. Recommendations on the choice of models are proposed.*

**Introduction**

Notwithstanding the efforts by the world software producers, the problem of the reduction of software insecurities has not yet been solved [7]. Objectively, the situation is due to the high structural complexity of the software systems, and dynamism of versions and technologies. One of the ways to increase the software security is the use of models at different stages of software testing to reliably evaluate the software security and the efficiency of its design. Most of such models are built on the basis of technology reliability theory, and that is why they are often called software reliability models [8,9]. From the point of view of the software reliability testing, the notion of software reliability is equivalent to the notion of software technological security, wherein a software error is understood as *a vulnerability*, that is, either code incorrectness, a defect, or an undeclared capabilities which can potentially affect the insecurity of the system and infrastructure [4].

The experience of testing laboratories shows that the use of the mathematical models must not distract the experts from the painstaking and responsible real software testing helping mainly at the decision making points. That is why it will be practical to classify the models in accordance with the input statistics at different life cycle stages of the software:

1. Test confidence models helping to estimate confidence in the software conformance evaluation;

2. Software complexity models helping to estimate the software complexity as well as related quality and security;

3. Reliability growth models helping to estimate software technological security depending on the testing time;

4. Debugging models helping to estimate software technological security depending on the input data runs in set data areas and following debugging processes.

**1. Test confidence models**

Test confidence models are based on independent introduction and following detection of test errors and independent expert examinations.

**1.1. Introduced Errors Consideration Model**

Introduced errors consideration model, also known as Labeled Fishes Problem of the probability theory, or the Mills model, is based on test errors introduced into the code. In testing, statistical data is collected on real and introduced errors revealed.

Test errors are supposed to be introduced stochastically, to make the identification of both own and introduced errors equally probable. In this case, we can estimate the original number of the software own errors by the use of maximum likelihood method:

$$N = \frac{Sn}{v},$$

where $S$ is the number of the mistakes introduced, $v$ is the number of identified introduced errors, and $n$ is the number of identified own errors.

The model can be used to estimate the accuracy of the code. Assuming that the testing is repeated until all introduced errors $S=v$ are found, the likelihood of the statement that the code contains $k$ own errors has the following expression:

$$R(k, S) = \begin{cases} 1, if\ n > k, \\ \frac{S}{S+k+1}, if\ n \leq k. \end{cases}$$

If the test does not identify all introduced test errors, the following formula must be used:

$$R(k, v) = \begin{cases} 1, n > k, \\ \frac{C_S^{v-1}n}{C_{S+k+1}^{k+v}}, n \leq k, \end{cases}$$

where $v$ is the number of identified test errors.

In practice, this model is used to control the efficiency of the experts who test the code security. A possible chance introduction of errors is considered one of the drawbacks of this model.

**1.2. Errors Introduction into Different Modules Model**

In this model, the software is divided into two parts. Assuming that Part 1 contains $N_1$ remaining errors and Part 2 $N_2$ remaining errors, the total number of the software remaining errors will be $N = N_1 + N_2$. It is assumed that the identification of the remaining errors is equally probable, and that one error is identified per time unit specified and corrected after identification.

It can be shown that:

$$\begin{cases} N_1(i) = N_1 - i + \sum_{j=0}^{i} \chi_j, \\ N_2(i) = N_2 - \sum_{j=0}^{i} \chi_j, \end{cases}$$

where $\chi_j$ is the characteristic function for which:

$$\chi_j = \begin{cases} 0, if\ the\ j^{th} error\ is\ found\ in\ Part\ 1\ of\ the\ software; \\ 1, if\ the\ j^{th} error\ is\ found\ in\ Part\ 2\ of\ the\ software. \end{cases}$$

In this case, the probability of the error identification in the set testing time interval can be evaluated $(t_i, t_{i+1}]$:

$$\begin{cases} p_1(i) = \frac{N_1(i)}{N_1(i)+N_2(i)} = \frac{N_1-i+\sum_{j=0}^{i}\chi_j}{N_1-i+N_2}, \\ p_2(i) = \frac{N_2(i)}{N_1(i)+N_2(i)} = \frac{N_2-\sum_{j=0}^{i}\chi_j}{N_1-i+N_2}. \end{cases}$$

$N_1$ and $N_2$ can be found with the maximum likelihood method.

### 1.3. Functional Objects Control Model

This model is used by the accredited laboratory [7] in testing software for undeclared opportunities. In accordance with the directive document by the Russian State Technical Committee, in the process of the software run-time analysis the set percentage $p$ of the $M_{fo}$ identified functional objects is being controlled. In the statistical analysis, $m_{fo}$ functional objects are arbitrary selected to introduce some test errors. In testing, test errors $s$ and own errors $n$ are identified. By analogy with the Basin approach [15], the software error number is evaluated with the use of the maximum likelihood method:

$$N = n \frac{M_{fo} - m_{fo} + 1}{\frac{p}{100} M_{fo} - s}.$$

### 1.4. Independent Expert Groups Testing Model

The group testing model presupposes testing by two independent testing or expert groups.

In testing, the numbers of errors found by either group, $N_1$ and $N_2$, are calculated, as well as the number of coinciding errors, $N_{12}$, found by both groups.

Granting that $N$ is the original number of errors, we can define the efficiency of either group as $E_1 = N_1/N$, and $E_2 = N_2/N$. If either group is hypothetically equally efficient, we may suppose that if Group 1 finds a certain number of errors out of the set number, Group 1 is able to define the same number of errors for any arbitrary selected subset. This yields:

$N_1/N = N_{12}/N_2$.

In this case, the intuitive model of the software original errors number evaluation will be:

$N = N_1 N_2 / N_{12}$.

This model is practicable when a simultaneous testing is performed by an independent expert group with their own test workbench, which is often the case in travelling tests under restricted time conditions.

### 2. Software Complexity Models

Software complexity models are based on the hypothesis that the software accuracy level can be predicted with the use of the software complexity metrics [19]. This will be true for non-deliberate insecurities, as the more complex and larger a code is, the higher is the probability that the programmer can make a mistake while writing or modifying it.

As a rule, software complexity metrics are used as the model arguments, while the models can be classed as a priori models and statistical models according to their complexity.

### 2.1. Halstead Metric Model of Errors

Among a priori software complexity model, the Halstead model is most commonly known. The model is based on two basic metrics of the software: directory $\eta$ of the operators $\eta_1$ and operands $\eta_2$ of the language and the number of uses of operators and operands $N$ in the code; also, the model is based on the supposition that the frequency of occurrence of operators and operands in the code is in direct proportion to the binary logarithm of the number of their types (by analogy with the information theory).

Basing on the general statistical physiological data of the human capabilities when writing a code possessing the above characteristics, a number of empirical models for software properties evaluation has been obtained.

The complexity of the code has been suggested to be understood as a number of intellectual actions (number of elementary problems solutions findings which a human being is capable to perform error-free) undertaken when writing a code in a certain language:

$$E = \widetilde{N} \log_2(\eta/L) = \frac{\eta_1 N_2 N \log_2 \eta}{2\eta_2},$$

where $\widetilde{N} = \eta_1 \log_2 \eta_1 + \eta_2 \log_2 \eta_2$ – is the theoretical code length, $\eta = \eta_1 + \eta_2$ - is the number of unique operators and operands of the language, $L = 2\eta_2/(\eta_1 N_2)$ - is the quality of the code, and $N = N_1 + N_2$ – is the number of uses of operators and operands in the software.

To calculate $D_0$ of the original number of errors, the following formula is suggested:

$$D_0 = \frac{V}{3000},$$

where $V = N \log_2(\eta_1 + \eta_2)$ - is the volume of the code (in bits).

### 2.2. Multifactor Complexity Model

The TRW phenomenological model [5] may be classed as a simple statistical model. This phenomenological model is a linear model of software complexity evaluation by the use of five empirical characteristics of the code, namely logical complexity $L_{tot}$, inter-linkage complexity $C_{inf}$, calculation complexity $C_c$, input-output complexity $C_{io}$, and readability $U_{read}$:

$$C = L_{tot} + 0.1 C_{inf} + 0.2 C_c + 0.4 C_{io} + (-0.1) U_{read}.$$

To calculate the number of errors $N$, the following multifactor model is suggested:

$$N = L_{tot}\kappa_1 + 0.1 C_{inf}\kappa_2 + 0.2 C_c \kappa_3 + 0.4 C_{io}\kappa_4 + (-0.1) U_{read}\kappa_5,$$

where $\kappa_i$ is the correlation index of the number of errors with the $i^{th}$ complexity metric.

The values of factors $\kappa_i$ can be easily found with the least squares method.

Among the disadvantages of the multifactor model, the "curse of dimensionality" is usually listed which becomes possible if a large number of factors is at work or if the polynomial approximant is non-linear.

### 3. Reliability Growth Models

Models of this class refer to the probabilistic dynamic models of discrete-state systems with continuous or discrete time. Such models are often referred to as time-domain reliability models. Popular models of this class are, for the most part, reducible to the homogeneous or nonhomogeneous Markov, or semi-Markov waiting line models [15].

The homogeneous Markov models presuppose that the total number of errors is an unknown finite constant. The number of errors remaining after testing and debugging is represented by an exponential law. The error rate $\lambda(i)$ depends on the current state of the system $i$ and does not depend on its previous states.

Unlike the homogeneous Markov models, the semi-Markov models presuppose that the error rate $\lambda(t_i)$ depends not only on the number of remaining errors, but also on the time $t_i$ that the system has been in this state.

Presently, among time-dependent models, the non-homogeneous Markov models are becoming more and more popular. In these models, the total number of errors is considered a random value described by the Poisson distribution, while the error rate is not a time-dependent linear function. This is why these models are often referred to as the Poisson models (Non-Homogeneous Poisson Process model, or NHPP models). Depending on the type of the Poisson intensity function, NHPP models are classed into convex (e. g., with Weibull and Pareto distribution), S-shaped (e. g., with Gompertz and Rayleigh distribution), and infinite (e. g., the logarithmical model).

There are modifications of the software reliability models done through the Bayesian method; these are sometimes referred to as a special type of models [20, 28].

It should be noted that for dynamic models calculations, the maximum likelihood method is traditionally employed, and rarely the linear regression methods and cross entropy methods.

Let us consider some examples of most popular time-dependent reliability growth models of each of the types.

**3.1. Exponential Reliability Growth Model**

Exponential reliability growth model known as the Jelinski-Moranda or JM model is based on the assumption that while software is being tested, the interval lengths between two consequential error identifications are distributed exponentially, hazard rate being in direct proportion to the number of unidentified errors. All errors are considered equally probable; each identified error is cured immediately, reducing the number of remaining errors by one.

Thus, the probability density function of the $i^{th}$ error identification time passed from the moment of the $(i – 1)^{th}$ error identification is as follows:

$$p(t_i) = \lambda_i e^{-\lambda_i t_i},$$

where $\lambda_i = \phi(N - (i - 1))$ - is the error rate being in direct proportion to the number of unidentified errors, $N$ is the original number of errors, $\phi$ - is the factor of proportionality understood as error detection rate, and $t_i$ - is the time interval between identifications of $(i-1)^{th}$ and $i^{th}$ errors.

The above formula permits to obtain formulae for error-free operation probability, complete debugging probability within a time given, mean time per one error identification, mean time of complete debugging, etc.

For calculation of $N$ and $\phi$ the maximum likelihood method is used.

Table 1 provides some examples of popular Markov models; $N$ stands for the original number of errors.

Table 1

**Markov Reliability Growth Models**

| Name | Error rate, $\lambda_i$ |
|---|---|
| JM model [18] | $\phi(N-(i-1))$ |
| Lipov-model [26] | $\phi(N - \sum_{j=1}^{i-1} N_j)$ |
| Xui model [30] | $\phi(e^{-k(N-i+1)} - 1)$ |
| Shanthikumar model [30] | $\phi(N-(i-1))^k$ |
| Bucchianico model [12] | $1 - \phi^{(N-(i-1))}$ |

### 3.2. The Rayleigh Reliability Growth Model

The Shick-Wolverton or SW model is a development of the JM model based on the assumption that the error rate is in direct proportion to not only the number of unidentified errors but also the debugging time interval length:

$$\lambda_i = \phi(N-(i-1))t_i,$$

where $N$ - is the original number of errors, $i$ is the number of errors identified, $\phi$ is the factor of proportionality understood as error detection rate, and $t_i$ is the time interval between identifications of $(i-1)^{th}$ and $i^{th}$ errors.

Thus, the Rayleigh distribution is deduced for the following probability density function:

$$p(t_i) = \phi(N-(i-1))t_i e^{-\phi(N-(i-1))\frac{t_i^2}{2}},$$

For calculation of $N$ and $\phi$ the maximum likelihood method is used.

Table 2 provides some examples of popular semi-Markov models; $N$ stands for the original number of errors.

Table 2

**Semi-Markov Reliability Growth Models**

| Name | Error rate, $\lambda_i$ |
|---|---|
| SW model [28] | $\phi(N-(i-1))\, t_i$ |
| Hyperbolic model [15] | $\phi(N-(i-1))(-a\, t_i^2 + bt_i + c)$ |
| Sukert model [30] | $\phi(N-(i-N_i))\, t_i$ |

| Modified Lipov model [26] | $\phi \left( N - \sum_{j=1}^{i-1} N_j \right) \left( \dfrac{t_i}{2} + \sum_{j=1}^{i-1} t_i \right)$ |

### 3.3. S-shaped NHPP Reliability Growth Model

At present, the S-shaped NHPP Yamada model is one of the most popular reliability growth models. The model assumes that the number of errors revealed per time unit is an independent random value of the Poisson distribution, the occurrence rate being in direct proportion to the expected number of remaining errors at the time given.

Unlike in JM- and SW-like convex models, this model includes an additional assumption that the number of errors has an S-shaped time dependence. Qualitatively, the S-shaped time dependence of the revealed errors is explained by the fact that at the initial stage of testing the tester spends some time on studying the software.

The number of errors function is as follows:

$m(t) = a(1 - (1 + gt)e^{-gt})$,

where $a$ is the factor of the expected errors, and $g$ is the factor of error detection rate.

Therefore, the error rate looks as follows:

$\lambda(t) = ag^2 t e^{-gt}$.

The above formulae permit to obtain formulae of probability of identifying (or non-identifying) a certain number of errors over time given.

The parameters of the model $a$ and $g$ are found through maximum likelihood method.

Examples of popular NHPP models are given in Table 3.

Table 3

**Non-homogenous Markov Reliability Growth Models**

| Name of an NHPP model | Number of errors function, $m(t)$ |
|---|---|
| Duane model [18] | $at^g$ |
| Gompertz model [14] | $ag^{c^t}$ |
| Goel-Okumoto model [30] | $a(1 - e^{-gt})$ |
| Schneidewind model [18] | $a/g(1 - e^{-gt})$ |
| Weibull model [16] | $a(1 - e^{-gt^c})$ |
| Exponential Yamada model [30] | $a(1 - e^{-r\,c(1-e^{(-gt)})})$ |
| Rayleigh S-shaped model [27] | $a(1 - e^{-r\,c(1-e^{(-\frac{gt^2}{2})})})$ |
| S-shaped delayed model [18] | $a(1 - (1 + gt)e^{-gt})$ |
| S-shaped model with flex point [20] | $\dfrac{a(1 - e^{-gt})}{1 + ce^{-gt}}$ |
| Parametrized S-curved model [15] | $a\dfrac{1 - e^{-gt}}{1 + \psi(r)e^{-gt}}$ |
| Dahiya model [16] | $a(1 - e^{-gt})/(1 + e^{-gt})$ |

| Pareto model [15] | $a(1 - (1 + t/c)^{1-g})$ |
| Hyperexponential model [20] | $a\left(1 - \sum_{i=1}^{2} b_i e^{-g_i t}\right)$ |
| Littlewood model [15] | $ac^g(\frac{1}{c^g} - \frac{1}{(c+t)^g})$ |
| Parabolical model [15] | $a(1 - e^{-((\frac{l}{3})t^3 + (\frac{m}{2})t^2 + nt)})$ |
| Logistic model [17] | $\frac{a}{1 + ke^{-gt}}$ |
| Pham model [27] | $a - ae^{-gt}(1 + (g+d)t + gdt^2)$ |
| Zhang model [14] | $\frac{a}{p-\beta}\left((1 - \frac{(1+\alpha)e^{-gt}}{1+\alpha e^{-gt}})^{\frac{c}{g}(p-\beta)}\right)$ |
| Xie logarithmical model [15] | $a\,ln^g(1+t)$ |
| Musa-Okumoto logarithmical model [18] | $1/a\,ln(agt+1)$ |

## 4. Software Debugging Models

Debugging models are based on the assumption that the software accuracy changes only with upgrades, and can be measured by software runs on set input data. Such models are often referred to as data-domain reliability models. Provisionally, debugging models may be classed into those based on input domain completeness those based on the level of upgrade.

### 4.1. The Nelson Structural Model

The model known as the Nelson model [5] is the Bernoulli binomial model with some rules added to regulate the use of input data.

In particular, the software input domain is defined as a set of $k$ non-overlapping domains $\{Z_i\}$ with single-valued correspondence to the set of probabilities $\{p_i\}$ of the event that the required input domain will be selected for the next software run. Therefore, if out of $N_i$ runs of the software (on $Z_i$ set of input data) $n_i$ runs ended in failure, the degree of the software reliability will be as follows:

$P = 1 - \sum_{i=1}^{k} \frac{n_i}{N_i} p_i$.

The model permits to calculate the probability $P_u$ of the software unfailing run when $n$ runs are performed:

$P_u = \prod_{j=1}^{u}(1 - Q_j) = e^{\left(\sum_{j=1}^{u} \ln(1-Q_j)\right)}$,

where $Q_j = \sum_{i=1}^{k} p_{ji}\chi_i$, $\chi_i$ - is a characteristic function of failure at the run on the $i^{th}$ input domain, and $p_{ji}$ is the probability of occurrence of the $i^{th}$ input domain in the $j^{th}$ run.

We must note that in the structural modification of the Nelson model the software graph analysis is suggested as the method of $p_{ji}$ finding. Unfortunately, it does not seem practicable for modifiable software of a complex structure. The need for a large number of tests to obtain exact evaluations is also considered a disadvantage of this type of model, but this difficulty is overcome by the use of the Wald statistical method [8].

The transition from debugging models to time-dependent models can be shown. Granting that $\Delta t_j$ is the time needed for the $j^{th}$ run yields the following:

$$P_u = e^{(\sum_{j=1}^{u} \lambda(t_j)\Delta t_j)},$$

where $\lambda(t_j) = \frac{-\ln(1-Q_j)}{\Delta t_j}$ is the failure rate, and $t_j = \sum_{i=1}^{j} \Delta t_i$ - is the total time needed for $j$ runs of the software.

Granting that $\Delta t_i$ is to become a relatively small value with the rise in the number of $u$ tests yields:

$$P(t) = e^{-\int_0^t \lambda(z)dz}.$$

### 4.2. Non-monotone Debugging and Upgrade Model

This model is based on the assumption that software reliability changes only with upgrades, and an upgrade can both raise or decrease the reliability:

$$P_u = P_0 + \sum_{j=1}^{u} \Delta P_j,$$

where $u$ is the number of upgrades, and $\Delta P_j$ is the delta of the reliability degree after the $j^{th}$ upgrade.

To consider the efficiency of the upgrade, $k_{ij}$ metric of the modified code is introduced for either debugging or software upgrades. This yields the basic formula of the software reliability [3]:

$$P_u = P_\infty - (P_\infty - P_0)\prod_{j=0}^{u}(1 - \sum_{i=1}^{2} a_i k_{ij}/P_\infty),$$

where $a_i$ is the factor of the efficiency of debugging or upgrade, $P_0$ is the original reliability degree, and $P_\infty$ is the ultimate reliability degree.

This model has four parameters ($P_0, P_\infty, a_1, a_2$) calculated with the maximum likelihood method. We must note that the model permits to yield formulae for test planning, for example, the number of remaining errors after the $u^{th}$ upgrade or debugging may be calculated as follows:

$$N_u = ceil\left(\frac{\ln(\frac{1-P_0}{1-P_u})}{\ln(1-a)}\right),$$

where $a$ is the average factor of the software reliability increase.

### 5. Selection of Evaluation Model and Planning of Testing

We must note that there is no universal model for software evaluation and testing planning. Moreover, apart from the models we have discussed, one can encounter descriptions of other types in the books, such as simulation models, structural models, fuzzy models, interval models, dynamic complexity models, software-hardware simulators, and neural nets; the latter are used for solving particular scientific problems. To choose a suitable model, we may suggest a number of qualitative and quantitative criteria.

Among qualitative criteria, we may name the following:

1. The simplicity of use. First of all, this relates to how the model answers the needs for statistical data collection. The used input data must be easily obtainable; they must be representative, and the input and output data must be comprehensible for the experts.

2. Validity, that is, the model must have a reasonable accuracy needed for solving the problems of analysis or synthesis in the software security domain. The positive quality of a model is its ability to make use of apriori information and complexing other models' data to reduce the input selection.

3. Usability for solving different problems. Some models permit to obtain an assessment of a wide range of factors that experts may need at different software lifecycle stages, for example, reliability indicators, expected rates of errors of different types, predictable time and cost expenditures, experts' proficiency, quality of tests, accuracy indicators, software overlapping indicators, etc.

4. Simplicity of implementation including the possibility to automate the assessment process basing on the existing mathematical packages and libraries, to retrain the model following upgrades, and to allow for incomplete or incorrect input statistics or for any other restriction.

The following quantitative criteria are also used:

- assessment accuracy indicators;

- predictive models quality indicators (precision, noise resilience, prediction accuracy, and coherence);

- information criteria of predictive models quality (dimensionality, and BIC/AIC criteria) [20].

- combined and integrated indicators, for example:

$$IC = \max \sum_{i=1}^{K} k_i \chi_i,$$

where $k_i$ is the weighting factor of the $i^{th}$ property of the model under consideration chosen by the expert; and $\chi_i$ is the characteristic function of the $i^{th}$ property.

## 4. Conclusions

Our research has revealed a great number of mathematical models that can be used to assess the technical software security at different stages of its lifecycle, which is very important for information security cost budgeting. The suggested classification of models will be practical when making the right choice or complexing models on the basis of available statistics.

One should bear in mind that because of rapid development, complexity, and diversity of modern software kits, the above models must not be expected ever to provide high accuracy, and very often they only provide intuitive data for taking a decision in preparation of software testing on the entire array of input data. Notwithstanding this, the results of these models applications are very convenient for use in both the justification of testing labour costs and reporting records, which may be helpful for the customer to view the obtained results as reliable.